\documentclass[11pt]{article}
\usepackage[T1]{fontenc}
\usepackage{amsmath,amssymb,amsthm}
\usepackage[hidelinks]{hyperref}

\title{Absolute monotonicity of Tops{\o}e's entropy ratio}
\author{Iosif Pinelis\\\small Michigan Technological University\\
\small\href{mailto:ipinelis@mtu.edu}{\texttt{ipinelis@mtu.edu}}}
\date{}

\newtheorem{theorem}{Theorem}
\newtheorem{lemma}{Lemma}

\newcommand\HH{\mathcal H}
\newcommand\la{\lambda}
\newcommand\de{\delta}
\newcommand\ep\varepsilon
\newcommand\be\beta
\newcommand\si\sigma

\newcommand\cc{\mathsf c}

\begin{document}
\maketitle

\begin{abstract}
For $p\in(0,1)$, let $H(p):=-p\ln p-(1-p)\ln(1-p)$ and consider
Tops{\o}e's entropy ratio
\begin{equation*}
R(p):=\frac{\ln\bigl(H(p)/\ln 2\bigr)}{\ln\bigl(4p(1-p)\bigr)},
\end{equation*}
with $R(1/2)$ interpreted by continuity.
We show that the coefficients in the expansion of $R(p)$ in powers
of $(1-2p)^2$ are strictly positive and form a Hausdorff moment sequence.
This implies the absolute monotonicity of $R(p)$ in
the variable $1-2p\in[0,1)$, as well as Tops{\o}e's conjectured
convexity of $R$. Our elementary proof uses two applications of an
integral quotient criterion and is an alternative to van de
Graaf's recent proof using a Pick-function representation.
\end{abstract}



For each $p\in(0,1)\setminus\{1/2\}$, let
\begin{equation*}
R(p):=\frac{\ln\bigl(H(p)/\ln 2\bigr)}{\ln\bigl(4p(1-p)\bigr)},
\end{equation*}
where $H(p):=-p\ln p-(1-p)\ln(1-p)$.
We extend the function $R$ to the entire interval $[0,1]$ by continuity.
For each
$x\in[-1,1]$, let also
\begin{equation*}
r(x):=R\Big(\frac{1-x}2\Big).
\end{equation*}

\begin{theorem}\label{thm:main}
There are strictly positive $d_0,d_1,\ldots$ such that
$\sum_{n\ge0}d_n=1$ and
\begin{equation}\label{eq:r(x)=}
r(x)=\sum_{n\ge0}d_n x^{2n}
\end{equation}
for 
$x\in[-1,1]$.
The sequence $(d_n)_{n\ge0}$ is a Hausdorff moment sequence
\cite{sokal2020}: there is a finite positive Borel measure
$\nu$ on $[0,1]$ such that
\begin{equation*}
d_n=\int_{[0,1]}t^n\,\nu(dt)
\end{equation*}
for every integer $n\ge0$.
In particular, $r$ is absolutely monotone on $[0,1)$, and
$r^{(k)}(x)>0$ for all integers $k\ge0$ and all $x\in(0,1)$.
Moreover, for all $p\in(0,1)$,
\begin{equation*}
R''(p)\ge R''(1/2)>0,
\end{equation*}
with equality $R''(p)=R''(1/2)$ only at $p=1/2$, so that $R$
is strictly convex on $[0,1]$.
\end{theorem}

Tops{\o}e \cite[Conjecture~6.2]{topsoe01} conjectured the convexity of
$R$ in 2001. In the same year, Pinelis \cite[Section~2.2]{pin01}
provided an algorithm for resolving this conjecture using l'Hospital-type
rules for monotonicity,
reducing the problem to an analysis
of the sign
pattern of a rational function; the
corresponding
computations were not carried out there. In 2026, van de Graaf
\cite[Theorem~1 and Corollary~7]{deGraaf} proved the convexity and the
strict positivity of the coefficients $d_n$ using analytic continuation
and a Pick-function integral representation with a positive density.
His proof includes an interval-arithmetic verification of one auxiliary
inequality, and his representation also implies that $(d_n)_{n\ge0}$
is a Hausdorff moment sequence. The present note gives an elementary
proof using the derivative ratio from \cite[Section~2.2]{pin01} and
two applications of an integral quotient criterion. We prove the
criterion by partial fractions and approximation; strict positivity
then follows from the moment representation and nonconstancy of its
generating function.

\section{A quotient lemma}

Let $\HH$ denote the class of functions $S$ defined by the formula
\begin{equation*}
S(s):=\int_{[0,1]}\frac{\eta(dt)}{1-st}
\end{equation*}
for 
$s\in[0,1)$,
where $\eta$ is a finite positive Borel measure. Expanding $\frac1{1-st}$ into the geometric
series shows that the functions in $\HH$ are exactly the generating functions of
Hausdorff moment sequences.
Let us refer to $\eta$ as the measure representing the function $S\in\HH$.

\begin{lemma}\label{lem:moment-quotient}
Let $\la$ be a
Borel measure on $[0,1]$, and
let $\phi$ and $\psi$ be
$\la$-integrable Borel
functions on $[0,1]$ with values in $[0,\infty)$ such that $\int_{[0,1]}\psi\,d\la>0$.
Let $Z:=\{t\in[0,1]\colon\phi(t)=\psi(t)=0\}$ and
$Z^\cc:=[0,1]\setminus Z$. Suppose that $\phi/\psi$ is nondecreasing
on $Z^\cc$, with the convention $u/0:=\infty$ for 
any 
$u\in(0,\infty)$.
For $s\in[0,1)$, let
\begin{equation}\label{eq:U,V}
U(s):=\int_{[0,1]}\frac{\phi(t)\,\la(dt)}{1-st}\quad\text{and}\quad
V(s):=\int_{[0,1]}\frac{\psi(t)\,\la(dt)}{1-st}.
\end{equation}
Then $Q:=
U/V\in\mathcal H$.
\end{lemma}

Unless stated otherwise, it is assumed in this note that $s$ is an arbitrary number in $[0,1)$.

Lemma~\ref{lem:moment-quotient} extends the quotient criterion of
Ruscheweyh, Salinas, and Sugawa
\cite[Theorem~1.10]{ruscheweyhSalinasSugawa} from the Lebesgue measure
to arbitrary Borel measures on $[0,1]$.
We give a real-variable proof of Lemma~\ref{lem:moment-quotient} by partial fractions and approximation, whereas
the proof in \cite{ruscheweyhSalinasSugawa} uses a characterization
by Pick functions.

\begin{proof}[Proof of Lemma~\ref{lem:moment-quotient}]
Replacing $\la$ by its restriction to $Z^\cc$ leaves $U$ and $V$
unchanged. We therefore may and will assume that $\la(Z)=0$.

We first consider the
case when $\la$ is a finite measure with a finite support $S_\la\subset(0,1]$ and $\psi>0$ on $S_\la$, so that we can write
\begin{equation}\label{eq:sums}
U(s)=\sum_{i=1}^m\frac{a_i}{1-t_i s}\quad\text{and}\quad
V(s)=\sum_{i=1}^m\frac{b_i}{1-t_i s},
\end{equation}
where $m$ is a positive integer,
$0<t_1<\cdots<t_m\le1$, $a_i\in[0,\infty)$ and $b_i\in(0,\infty)$ for all $i\in[m]:=\{1,\ldots,m\}$, and
$k_i:=a_i/b_i$ is nondecreasing in $i\in[m]$.

For 
all real
$s\notin\{1/t_1,\ldots,1/t_m\}$,
$
V'(s)=\sum_{i=1}^m\frac{b_it_i}{(1-t_i s)^2}>0.
$
For each $i\in[m-1]$, the function $V$ increases from
$-\infty$ to $+\infty$ on $(1/t_{i+1},1/t_i)$ and thus has a simple zero  $z_i\in(1/t_{i+1},1/t_i)\subset(1,\infty)$.
These
$m-1$ simple zeros, $z_1,\dots,z_{m-1}$, are all the zeros of $V$, since the numerator of the rational function $V$ after taking a
common denominator is of degree $\le m-1$.
For any $j\in[m-1]$, we have $t_j<1/z_j<t_{j+1}$ and hence
\begin{equation*}
U(z_j)=U(z_j)-k_j V(z_j)
=\sum_{i=1}^m\frac{b_i(k_i-k_j)}{1-t_i z_j}
=-\sum_{i=1}^m\frac{b_i}{z_j}\frac{k_i-k_j}{t_i-1/z_j}\le0,
\end{equation*}
because $k_i$ is nondecreasing in $i\in[m]$ and hence
$\frac{k_i-k_j}{t_i-1/z_j}\ge0$ for all $i\in[m]$.

The singularities of $U/V$ at the poles $1/t_i$ are removable. So, we have a partial fractions decomposition
\begin{equation*}
\frac{U(s)}{V(s)}
=c_0+\sum_{j=1}^{m-1}\frac{c_j}{1-s/z_j},
\end{equation*}
where
\begin{equation*}
c_0=\lim_{s\to-\infty}\frac{U(s)}{V(s)}
=\frac{\sum_{i=1}^m a_i/t_i}{\sum_{i=1}^m b_i/t_i}\ge0
\end{equation*}
and, for 
all $j\in[m-1]$,
\begin{equation*}
c_j=-\frac{U(z_j)}{z_jV'(z_j)}\ge0.
\end{equation*}
Thus, $U/V\in\HH$, with representing measure
$c_0\de_0+\sum_{j=1}^{m-1}c_j\de_{1/z_j}$, where $\de_y$ is the
Dirac probability measure supported on the singleton set $\{y\}$.

We now approximate the integrals
in \eqref{eq:U,V} by sums as in \eqref{eq:sums}.
For each integer $N\ge1$, let
$I_{N,i}:=[(i-1)/N,i/N)$ for $i\in[N-1]$ and
$I_{N,N}:=[(N-1)/N,1]$. For each $i\in[N]$, let
\begin{equation*}
t_{N,i}:=\frac{i}{N},\qquad
a_{N,i}:=\int_{I_{N,i}}\phi\,d\la,\qquad
b_{N,i}:=\int_{I_{N,i}}\psi\,d\la, \quad
J_{N,i}:=I_{N,i}\cap Z^\cc.
\end{equation*}
Let also $\psi_N:=\psi+\phi/N$. Then $\phi/\psi_N\in[0,N]$ on $Z^\cc$. The condition that $\phi/\psi$ is nondecreasing
on $Z^\cc$ implies that, with the standard conventions $\sup\emptyset:=-\infty$ and $\inf\emptyset:=\infty$, for each $i\in[N-1]$ there is some $K_{N,i}\in[0,N]$ such that
\begin{equation*}
	\sup_{J_{N,i}}\frac\phi{\psi_N}\le K_{N,i}\le\inf_{J_{N,i+1}}\frac\phi{\psi_N}.
\end{equation*}
So, letting $\be_{N,i}:=b_{N,i}+a_{N,i}/N$ for $i\in[N]$, for each $i\in[N-1]$ we have
\begin{equation*}
	a_{N,i}=\int_{I_{N,i}}\phi\,d\la=\int_{J_{N,i}}\phi\,d\la
	\le K_{N,i}\int_{J_{N,i}}\psi_N\,d\la=K_{N,i}\be_{N,i}
\end{equation*}
and, similarly, $a_{N,i+1}\ge K_{N,i}\be_{N,i+1}$.
Letting now $Z_N:=\{i\in[N]\colon \be_{N,i}=0\}$ and $Z_N^\cc:=[N]\setminus Z_N$, we see that for some nondecreasing family $(k_{N,i})_{i\in Z_N^\cc}$ in $[0,N]$ we have $a_{N,i}=k_{N,i}\be_{N,i}$ for all $i\in Z_N^\cc$.

Let $\tau_N(t):=t_{N,i}$ for $i\in[N]$ and $t\in I_{N,i}$. Then
\begin{equation*}
\begin{aligned}
U_N(s)&:=\sum_{i\in Z_N^\cc}\frac{a_{N,i}}{1-t_{N,i}s}
=\sum_{i=1}^N\frac{a_{N,i}}{1-t_{N,i}s}
=\int_{[0,1]}\frac{\phi(t)\,\la(dt)}{1-s\tau_N(t)},\\
V_N(s)&:=\sum_{i\in Z_N^\cc}\frac{b_{N,i}}{1-t_{N,i}s}
=\sum_{i=1}^N\frac{b_{N,i}}{1-t_{N,i}s}
=\int_{[0,1]}\frac{\psi(t)\,\la(dt)}{1-s\tau_N(t)}.
\end{aligned}
\end{equation*}
Since $\tau_N(t)\ge t$ for 
all $t\in[0,1]$, we have
$V_N(s)\ge V(s)>0$ for 
all $s\in[0,1)$.
Applying the finite-sum result to the
weights $a_{N,i}$ and $\be_{N,i}$ with $i\in Z_N^\cc$ instead of $a_i$ and $b_i$ with $i\in[m]$ in \eqref{eq:sums} gives
\begin{equation*}
Q_N:=\frac{U_N}{V_N+U_N/N}\in\HH.
\end{equation*}

For each $s\in[0,1)$, we have
\begin{equation*}
\frac1{1-s\tau_N(t)}\longrightarrow\frac1{1-st}
\end{equation*}
uniformly in $t\in[0,1]$ as $N\to\infty$.
Thus, $U_N(s)\to U(s)$ and $V_N(s)\to V(s)>0$ as $N\to\infty$,
and hence $Q_N(s)\to Q(s)$.
Let $\eta_N$ be a measure representing $Q_N$. Its total mass is
\begin{equation*}
\eta_N([0,1])=Q_N(0)
=\frac{U(0)}{V(0)+U(0)/N}\le\frac{U(0)}{V(0)}<\infty.
\end{equation*}
By weak compactness, there are a subsequence $(\eta_{N_\ell})$
and a finite positive measure $\eta$ on $[0,1]$ such that
$\eta_{N_\ell}$ converges weakly to $\eta$ as $\ell\to\infty$.
For every $s\in[0,1)$, the function $t\mapsto(1-st)^{-1}$ is
continuous on $[0,1]$, so
\begin{equation*}
Q(s)=\lim_{\ell\to\infty}Q_{N_\ell}(s)
=\lim_{\ell\to\infty}\int_{[0,1]}\frac{\eta_{N_\ell}(dt)}{1-st}
=\int_{[0,1]}\frac{\eta(dt)}{1-st}.
\end{equation*}
This proves $Q\in\HH$.
\end{proof}

\paragraph{Remark.}
A related criterion for Laplace transforms was obtained by Mao and
Tian \cite[Main Theorem]{maoTian}. Let $g>0$ be log convex on
$(0,\infty)$, let $f\ge0$ be Borel, and suppose both Laplace
transforms are finite on $(0,\infty)$. If $f/g$ is nondecreasing,
then $(\mathcal Lf)/(\mathcal Lg)$ is completely monotone, where
$\mathcal L$ denotes the Laplace transform.

\section{The moment representation}\label{sec:moments}

Let $h(x):=H((1-x)/2)$ for $x\in(-1,1)$. In this section
and the next, let $s\in(0,1)$ unless stated otherwise, and let
\begin{equation*}
F(s):=-\ln\bigl(h(\sqrt s)/\ln 2\bigr),\qquad G(s):=-\ln(1-s).
\end{equation*}
Then $F(0+)=G(0+)=0$. Let $D:=F/G$. By the symmetry of $h$,
we have $D(x^2)=r(x)$ for $x\in(-1,1)\setminus\{0\}$.

\paragraph{The derivative ratio and the first application.}
For 
all $x\in(-1,1)$, we have
\begin{equation*}
h'(x)=-xA(x^2)\quad\text{and}\quad h(x)=(1-x^2)B(x^2),
\end{equation*}
where 
\begin{equation*}
A(s):=\int_0^1\frac{dt}{1-st^2}\quad\text{and}\quad
B(s):=\int_0^1\frac{dt}{(1+t)(1-st^2)} 
\end{equation*}
for $s\in[0,1)$. 

Let $C:=A/B$. Then, for $s\in(0,1)$,
\begin{equation}\label{eq:derivatives}
F'(s)=\frac{C(s)}{2(1-s)}.
\end{equation}
Using the substitution $u=t^2$ and letting $\la(du):=\frac{du}{2\sqrt u(1+\sqrt u)}$ on $(0,1)$, we have
\begin{equation*}
A(s)=\int_0^1\frac{1+\sqrt u}{1-su}\,\la(du)\quad\text{and}\quad
B(s)=\int_0^1\frac{\la(du)}{1-su}.
\end{equation*}
Now Lemma~\ref{lem:moment-quotient}, with $\phi(u)=1+\sqrt u$ and
$\psi=1$, gives a finite positive measure $\mu$ on $[0,1]$ such that
\begin{equation}\label{eq:C-moments}
C(s)=\int_{[0,1]}\frac{\mu(dt)}{1-st}
\end{equation}
for 
$s\in[0,1)$.
Thus, the Taylor coefficients of $C$
form a Hausdorff moment
sequence. Clearly, $A(s)<2B(s)$, hence
$C(s)<2$ for $s\in[0,1)$. By monotone convergence in
\eqref{eq:C-moments},
\begin{equation}\label{eq:bounded}
\mu(\{1\})=0\quad\text{and}\quad
\int_{[0,1)}\frac{\mu(dt)}{1-t}\le2.
\end{equation}

\paragraph{Integration and the second application.}
For 
all $t\in[0,1)$ and $s\in(0,1)$, 
\begin{equation*}
\frac1s\int_0^s\frac{d\si}{(1-\si)(1-\si t)}
=\frac{\ln(1-st)-\ln(1-s)}{s (1-t)}
=\frac1{1-t}\int_t^1\frac{du}{1-su}.
\end{equation*}
Using $F(0+)=0$, \eqref{eq:derivatives}, \eqref{eq:C-moments},
and Tonelli's theorem, we therefore obtain
\begin{equation*}
\frac{F(s)}s
=\frac12\int_{[0,1)}\frac{\mu(dt)}{1-t}
  \int_t^1\frac{du}{1-su}
=\int_0^1\frac{q(u)}{1-su}\,du,
\end{equation*}
where
\begin{equation*}
q(u):=\frac12\int_{[0,u]}\frac{\mu(dt)}{1-t}
\end{equation*}
for $u\in[0,1]$.
The function $q$ is nonnegative, nondecreasing, and bounded by~$1$, in view of \eqref{eq:bounded}.
Also,
\begin{equation*}
\frac{G(s)}s=\int_0^1\frac{du}{1-su}.
\end{equation*}
Applying Lemma~\ref{lem:moment-quotient} with Lebesgue measure,
$\phi=q$, and $\psi=1$ gives a finite positive measure $\nu$
on $[0,1]$ such that
\begin{equation}\label{eq:D-moments}
D(s)=\frac{F(s)}{G(s)}=\int_{[0,1]}\frac{\nu(dt)}{1-st}
=\sum_{n\ge0}d_n s^n, 
\quad\text{where}\quad 
d_n:=\int_{[0,1]}t^n\,\nu(dt)\ge0
\end{equation}
for 
integers $n\ge0$.
The series converges absolutely for every $s\in(-1,1)$.
Consequently, extending $D$ to zero by continuity, we have
\begin{equation}\label{eq:r=series}
r(x)=D(x^2)=\sum_{n\ge0}d_n x^{2n}
\end{equation}
for 
$x\in(-1,1)$.

\section{Strict positivity and consequences}

Since $C(s)<2$, \eqref{eq:derivatives} gives $F'(s)<G'(s)$.
Together with $F(0+)=G(0+)=0$, this implies $D(s)<1$ for
$s\in(0,1)$. On the other hand,
\begin{equation*}
D(1-)=R(0+)=1,
\end{equation*}
because $H(p)\sim-p\ln p$ as $p\downarrow0$.
Thus $D$ is nonconstant. By \eqref{eq:D-moments}, the measure
$\nu$ cannot be supported on $\{0\}$, so that $\nu((0,1])>0$.
It follows that
\begin{equation*}
d_n=\int_{[0,1]}t^n\,\nu(dt)>0
\end{equation*}
for 
integers $n\ge0$.

Letting $s\uparrow1$ in \eqref{eq:D-moments} and using monotone
convergence gives $\sum_{n\ge0}d_n=1$. Therefore the series for $r$
converges uniformly on $[-1,1]$ and represents $r$ also at
$x=\pm1$. This, together with \eqref{eq:r=series}, completes the proof of \eqref{eq:r(x)=}.

Termwise differentiation in \eqref{eq:r=series} gives $r^{(k)}(x)\ge0$ for all
integers $k\ge0$ and all $x\in[0,1)$, with strict inequality for
$x\in(0,1)$. Finally, for 
all $p\in(0,1)$,
\begin{equation*}
R''(p)=4r''(1-2p)\ge4r''(0)=8d_1>0.
\end{equation*}
Since $d_2>0$, the first inequality is strict when $p\ne1/2$.
Together with continuity at the endpoints, this also proves strict
convexity on $[0,1]$ and completes the proof of
Theorem~\ref{thm:main}.

\paragraph{Use of AI tools.}
This note was developed through an iterative collaboration between
Iosif Pinelis and OpenAI's Codex assistant, referred to here as Theo.
Theo developed most of the mathematical argument and drafted most of
the text, and assisted with checking calculations, locating and verifying
references, and preparing the \LaTeX{} source. Pinelis selected the
problem, provided its mathematical and historical context, and directed
the investigation. He suggested seeking an elementary proof of the
Hausdorff moment property through moment generating functions and their
quotients, and using that property to establish strict positivity,
thereby eliminating the separate coefficient argument. He reviewed and
revised the mathematical arguments and exposition and takes full
responsibility for the final manuscript, including the correctness of
the proofs and the accuracy and completeness of the attribution to
prior work.

\bibliographystyle{abbrv}
\begingroup
\interlinepenalty=10000
\bibliography{arxiv_v2-references}

@article{sokal2020,
  author  = {Sokal, Alan D.},
  title   = {{W}all's continued-fraction characterization of {H}ausdorff
             moment sequences: A conceptual proof},
  journal = {Proc. Amer. Math. Soc.},
  volume  = {148},
  number  = {5},
  year    = {2020},
  pages   = {2111--2116},
  doi     = {10.1090/proc/14884},
  note    = {DOI: \href{https://doi.org/10.1090/proc/14884}{\nolinkurl{10.1090/proc/14884}}},
  url     = {https://discovery.ucl.ac.uk/id/eprint/10082936/1/wall-hausdorff_v2.pdf}
}

@article{maoTian,
  author  = {Mao, Zhong-Xuan and Tian, Jing-Feng},
  title   = {Complete monotonicity of {L}aplace transform ratios:
             Density criteria for self-decomposability and additive decomposition},
  journal = {Preprint},
  year    = {2026},
  note    = {\href{https://arxiv.org/abs/2611.24348}{arXiv:2611.24348}},
  url     = {https://arxiv.org/abs/2611.24348}
}

@article {pin01,
    AUTHOR = {Pinelis, Iosif},
     TITLE = {{L}'{H}ospital type rules for oscillation, with applications},
   JOURNAL = {JIPAM. J. Inequal. Pure Appl. Math.},
  FJOURNAL = {JIPAM. Journal of Inequalities in Pure and Applied
              Mathematics},
    VOLUME = {2},
      YEAR = {2001},
    NUMBER = {3},
     PAGES = {Article 33, 24 pp. (electronic)},
      ISSN = {1443-5756},
   MRCLASS = {26B25 (60E15 62P10)},
  MRNUMBER = {MR1876266 (2002j:26008)},
MRREVIEWER = {Winston T. Lin},
}

@article{topsoe01,
  author  = {Tops{\o}e, Flemming},
  title   = {Bounds for entropy and divergence for distributions over a two-element set},
  journal = {JIPAM. J. Inequal. Pure Appl. Math.},
  volume  = {2},
  number  = {2},
  year    = {2001},
  pages   = {Article 25, 13 pp. (electronic)},
  url     = {https://www.emis.de/journals/JIPAM/images/044_00_JIPAM/044_00.pdf}
}

@article{deGraaf,
    title={Why $(4x(1-x))^{3/4}$ approximates the binary entropy
so well},
    journal={Preprint},
    author={van de Graaf, Jeroen},
    year={2026},
    note={\href{https://arxiv.org/abs/2611.33213}{arXiv:2611.33213}},
}

@article{ruscheweyhSalinasSugawa,
  author  = {Ruscheweyh, Stephan and Salinas, Luis and Sugawa, Toshiyuki},
  title   = {Completely monotone sequences and universally prestarlike functions},
  journal = {Israel J. Math.},
  volume  = {171},
  year    = {2009},
  pages   = {285--304},
  doi     = {10.1007/s11856-009-0050-9},
  note    = {DOI: \href{https://doi.org/10.1007/s11856-009-0050-9}{\nolinkurl{10.1007/s11856-009-0050-9}}},
  url     = {https://doi.org/10.1007/s11856-009-0050-9}
}
\endgroup

\end{document}